\documentclass[aps,prd,twocolumn,showpacs,amsmath,amssymb]{revtex4-1}
\usepackage{amsmath}
\usepackage{graphicx}
\usepackage{subfigure}
\usepackage{epstopdf}
\usepackage{color}
\usepackage{multirow}
\usepackage{setspace}
\usepackage{overpic}
\usepackage{amssymb}
\usepackage[bookmarksnumbered, pdfstartview=FitH,colorlinks,urlcolor=blue, citecolor=blue,linkcolor=blue] {hyperref}
\usepackage{lineno}
\usepackage{bm}
\usepackage{rotating}
\usepackage[utf8]{inputenc}
\let\oldequation\equation
\let\oldendequation\endequation

\renewenvironment{equation}
  {\linenomathNonumbers\oldequation}
  {\oldendequation\endlinenomath}

\begin{document}
%\linenumbers

\title{\bf \boldmath
Search for baryon-number and lepton-number violating decays $D^{0}\to \bar{p}e^{+}$ and $D^{0}\to pe^{-}$}

\author{
%\begin{small}
%\begin{center}
M.~Ablikim$^{1}$, M.~N.~Achasov$^{10,b}$, P.~Adlarson$^{67}$, S. ~Ahmed$^{15}$, M.~Albrecht$^{4}$, R.~Aliberti$^{28}$, A.~Amoroso$^{66A,66C}$, M.~R.~An$^{32}$, Q.~An$^{63,49}$, X.~H.~Bai$^{57}$
% ...
\\
\vspace{0.2cm}
(BESIII Collaboration)\\
\vspace{0.2cm} {\it
$^{1}$ Institute of High Energy Physics, Beijing 100049, People's Republic of China\\
% ...
$^{4}$ Bochum Ruhr-University, D-44780 Bochum, Germany\\
% ...
$^{10}$ G.I. Budker Institute of Nuclear Physics SB RAS (BINP), Novosibirsk 630090, Russia\\
% ...
$^{15}$ Helmholtz Institute Mainz, Staudinger Weg 18, D-55099 Mainz, Germany\\
% ...
$^{28}$ Johannes Gutenberg University of Mainz, Johann-Joachim-Becher-Weg 45, D-55099 Mainz, Germany\\
% ...
$^{32}$ Liaoning Normal University, Dalian 116029, People's Republic of China\\
% ...
$^{49}$ State Key Laboratory of Particle Detection and Electronics, Beijing 100049, Hefei 230026, People's Republic of China\\
% ...
$^{57}$ University of Jinan, Jinan 250022, People's Republic of China\\
% ...
$^{63}$ University of Science and Technology of China, Hefei 230026, People's Republic of China\\
% ...
$^{66}$ University of Turin and INFN, (A)University of Turin, I-10125, Turin, Italy; (B)University of Eastern Piedmont, I-15121, Alessandria, Italy; (C)INFN, I-10125, Turin, Italy\\
$^{67}$ Uppsala University, Box 516, SE-75120 Uppsala, Sweden\\
% ...
$^{b}$ Also at the Novosibirsk State University, Novosibirsk, 630090, Russia\\
% ...
}
%\end{center}
%\end{small}
}

\begin{abstract}
Using an electron-positron collision data sample corresponding to an integrated luminosity of 2.93~fb$^{-1}$ collected with the BESIII detector at a center-of-mass energy of 3.773~GeV, we search for the baryon-number and lepton-number violating decays $D^{0}\to \bar{p}e^{+}$ and $D^{0}\to pe^{-}$.
No obvious signals are found with the current statistics. The upper limits on the branching fractions for
$D^{0}\to \bar{p}e^{+}$ and $D^{0}\to pe^{-}$ are set to be $1.2\times 10^{-6}$ and $2.2\times 10^{-6}$ at 90\% confidence level, respectively.
\end{abstract}


\maketitle

\oddsidemargin  -0.2cm
\evensidemargin -0.2cm

\section{Introduction}

As demonstrated by the stability of ordinary matter, baryon number ($B$) is empirically known to be conserved to a very high degree. However, the absolute conservation of $B$ has been questioned for many years.
For example, the fact that there is an excess of baryons over anti-baryons in the Universe implies the existence of baryon number violating~(BNV) processes. Therefore, various extensions of the Standard Model (SM) with BNV processes have been proposed.
At the level of dimension-six operators, BNV processes can happen with $\Delta(B-L)=0$, where~$\Delta(B-L)$ is the change of baryon number minus lepton number between initial and final states~\cite{B1}. Another class of BNV operators are the dimension-seven operators allowing $\Delta(B-L)=2$ processes~\cite{B2}.
Some of the SM extensions, \emph{e.g.}, $\rm{SU}(5)$, $\rm{SO}(10)$, $\rm{E}6$ and flipped $\rm{SU}(5)$ models, predict branching fractions~(BFs) for these kinds of decays at the level of $10^{-39}$ to $10^{-27}$~\cite{su5,pre}, compatible with the experimental limits from proton decay experiments.

\begin{figure}[htbp]
\centering
\includegraphics[height=2.6cm]{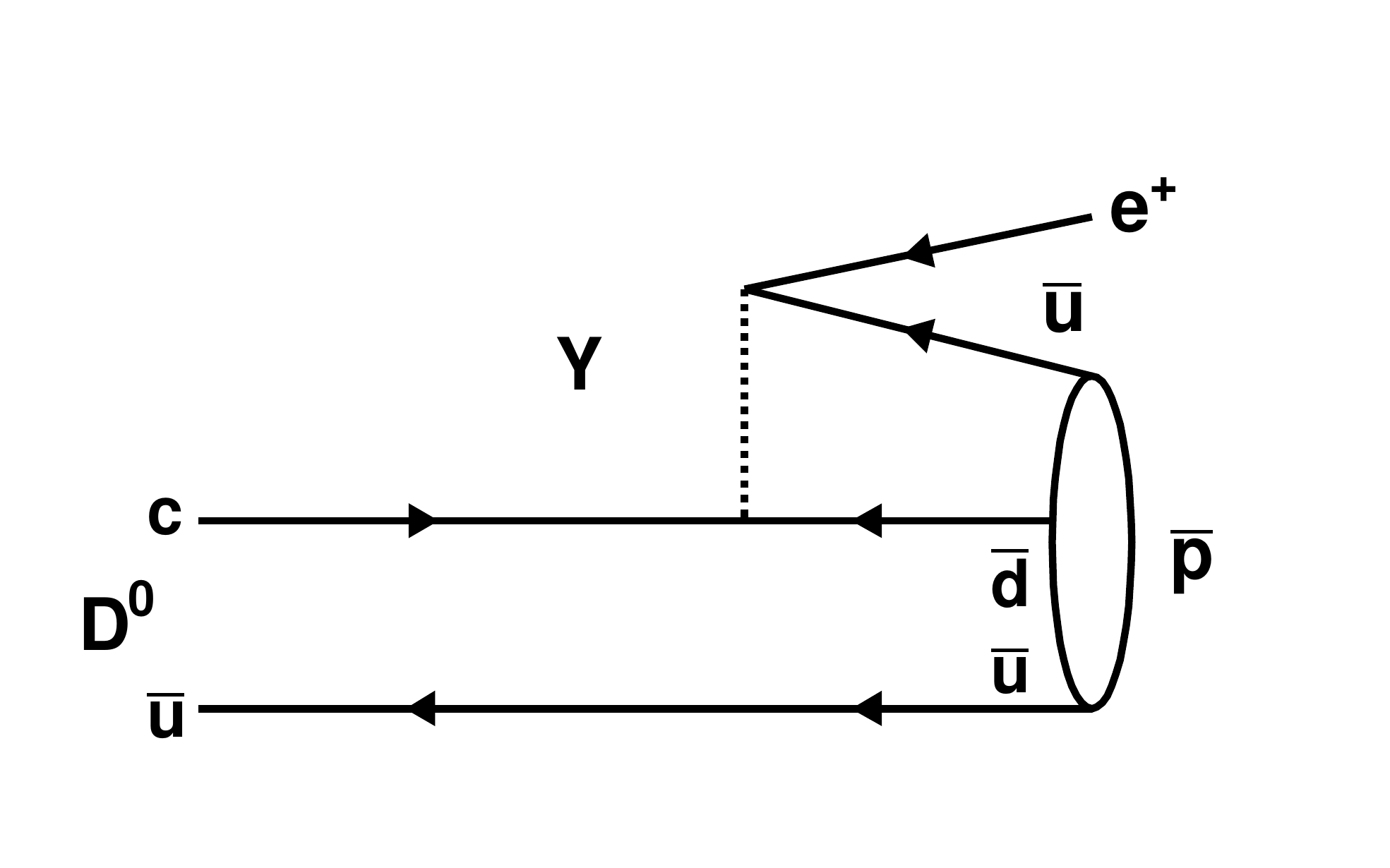}
\includegraphics[height=2.6cm]{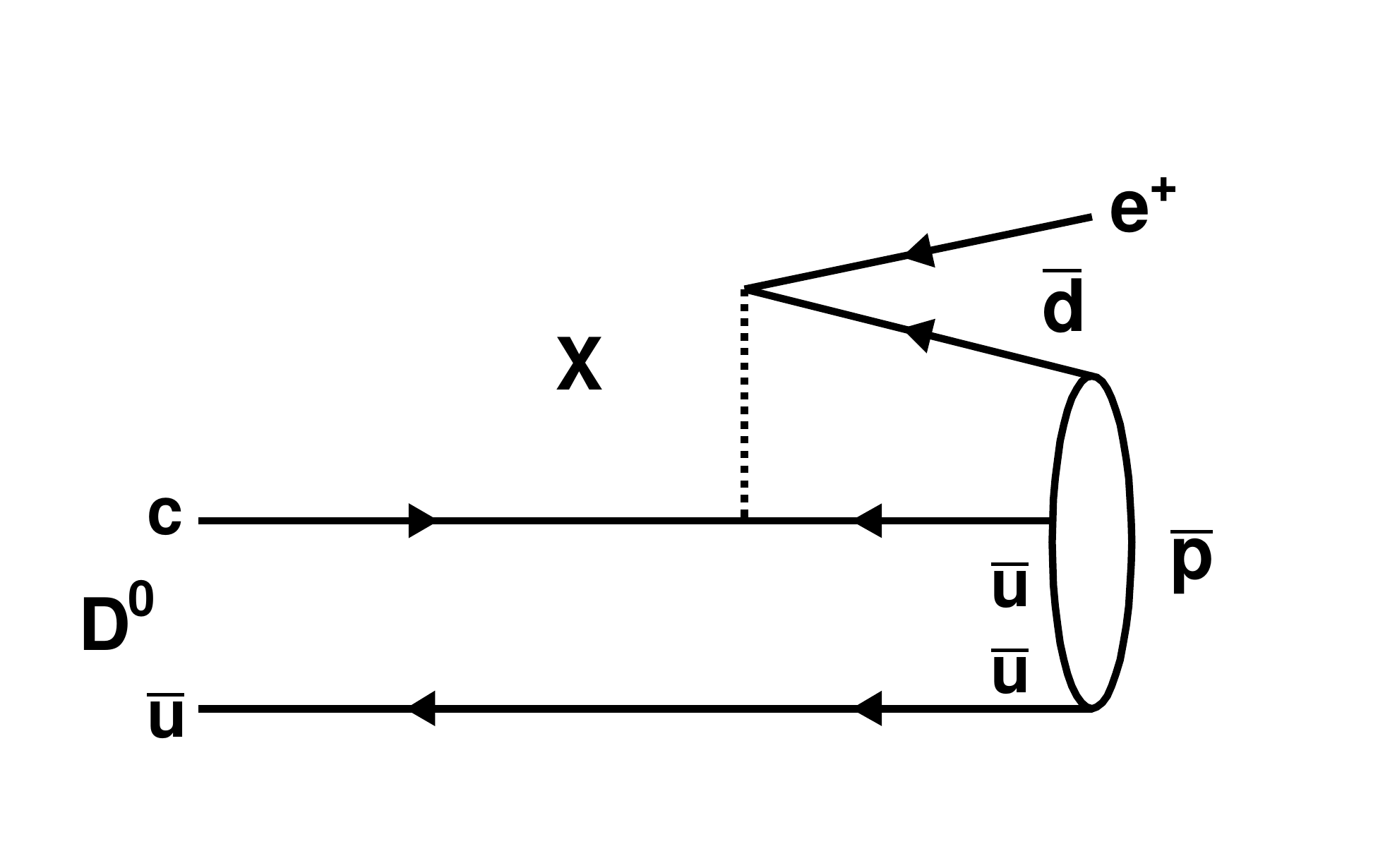}
\caption{Feynman diagrams of $D^{0}\to \bar p e^{+}$ based on a leptoquark scenario.}
\label{fig::cleo}
\end{figure}

For decades, the decay of the proton, the lightest baryon, has been searched for without success. An alternative probe is to look for the BNV decays of a heavy quark.
In 2009, the CLEO collaboration searched for the decays of $D^0(\bar{D}^0)\to \bar pe^+$ and $D^0(\bar{D}^0)\to pe^-$~\cite{cleo} and set upper limits~(ULs) on the BFs to be $\mathcal{B}(D^0(\bar{D}^0)\to \bar pe^+)<1.1\times 10^{-5}$ and $\mathcal{B}(D^0(\bar{D}^0)\to pe^-)<1.0\times 10^{-5}$ at 90\% confidence level~(CL), respectively. For this result, the initial flavor ($D^0$ vs.~$\bar{D}^0$) of the charm meson was not determined.  The Feyman diagrams in Fig.~\ref{fig::cleo}~\cite{cleo} show some of the possible mechanisms of $D^{0}\to \bar p e^{+}$ based on analogous couplings of $p\to e^+\pi^0$ in SU(5) which is suggested by Biswal {\it et al}~\cite{cleo,semilepton}. However, there is no tree-level diagram for $D^0 \to pe^-$ in SU(5).
These decays can be mediated by heavy hypothetical gauge bosons $X$ and $Y$ which have electric charge $\frac{4}{3}e$ or $\frac{1}{3}e$ and can couple a quark to a lepton. Hence, these bosons are sometimes called ``leptoquarks''.
Various BNV processes were searched for in $\tau$, $\Lambda$, $D$ and $B$ decays by the CLEO~\cite{cleo2}, CLAS~\cite{clas} and BaBar~\cite{babar} experiments, but no evidence was found.
The large data samples accumulated by the BESIII experiment lead to the best sensitivity for investigating BNV decays of charmed mesons or charmonium states.  The BESIII collaboration searched for BNV in $D^{+}\to \bar{\Lambda}(\bar{\Sigma}^0)e^{+}$~\cite{BNV1} and $J/\psi \to \Lambda_{c}^{+}e^{-} + c.c$~\cite{BNV2} and set ULs at the level of $10^{-8}-10^{-6}$ with no significant signals.

In this paper, we present the most accurate search to date for the decays $D^0\to pe^-$ and $D^0 \to \bar p e^+$ performed with an $e^+e^-$ collision data sample corresponding to an integrated luminosity of 2.93~fb$^{-1}$~\cite{lum_bes3} taken at a center-of-mass~(CM) energy of 3.773~GeV with the BESIII detector. Throughout this paper, charge conjugate channels are always implied.

\section{BESIII detector and Monte Carlo simulation}

The BESIII detector is a magnetic
spectrometer~\cite{BESIII} located at the Beijing Electron
Positron Collider (BEPCII)~\cite{Yu:IPAC2016-TUYA01}. The
cylindrical core of the BESIII detector consists of a helium-based
 multilayer drift chamber (MDC), a plastic scintillator time-of-flight
system (TOF), and a CsI(Tl) electromagnetic calorimeter (EMC),
which are all enclosed in a superconducting solenoidal magnet
providing a 1.0~T magnetic field. The solenoid is supported by an
octagonal flux-return yoke with resistive plate counter muon
identifier modules interleaved with steel. The acceptance of
charged particles and photons is 93\% over $4\pi$ solid angle. The
charged-particle momentum resolution at $1~{\rm GeV}/c$ is
$0.5\%$, and the specific energy loss ($dE/dx$) resolution is $6\%$ for the electrons
from Bhabha scattering. The EMC measures photon energies with a
resolution of $2.5\%$ ($5\%$) at $1$~GeV in the barrel (end cap)
region. The time resolution of the TOF barrel part is 68~ps, while
that of the end cap part is 110~ps.

Monte Carlo (MC) simulated  data samples produced with a {\sc geant4}-based~\cite{geant4} package, which
includes the geometric description of the BESIII detector and the
detector response, are used to determine the detection efficiency
and to estimate the backgrounds. The simulation includes the beam
energy spread and initial state radiation (ISR) in the $e^+e^-$
annihilations modeled with the generator {\sc kkmc}~\cite{kkmc}.
The inclusive MC samples consist of the production of $D\bar{D}$
pairs with consideration of quantum coherence for all neutral $D$
modes, the non-$D\bar{D}$ decays of the $\psi(3770)$, the ISR
production of the $J/\psi$ and $\psi(3686)$ states, and the
continuum processes.
The known decay modes are modeled with {\sc
evtgen}~\cite{evtgen} using the BFs taken from the
Particle Data Group~\cite{pdg2018}, and the remaining unknown decays
from the charmonium states with {\sc
lundcharm}~\cite{lundcharm}. The final state radiations
from charged final state~(FSR) particles are incorporated with the {\sc
photos} package~\cite{photos}.

\section{Data analysis}
\subsection{Method}

At $\sqrt s=3.773$ GeV, $D^0\bar D^0$ meson pairs are produced from $\psi(3770)$ decays without accompanying hadron(s). This offers an ideal platform to investigate the rare decays of $D^{0}$ in a very low background environment by using the double-tag~(DT) method~\cite{mark3}.

An event where a $\bar D^0$ is reconstructed via the hadronic decay modes of
$\bar D^0\to K^+\pi^-$, $K^+\pi^-\pi^0$ or $K^+\pi^-\pi^-\pi^+$ is called a single-tag~(ST) candidate event.
For fully-reconstructed STs, the remaining tracks and showers originate from the other meson, the $D^0$.
An event in which the decays of the $D^0$ and $\bar D^0$ are both reconstructed is called a DT candidate event.
In this work, we search for the events with $D^0$ decays into $\bar{p}e^+$ or $pe^-$ and $\bar D^0$ decays into one of the above three hadronic channels.
The BFs for $D^0\to \bar{p} e^{+}$ and $D^{0}\to pe^{-}$ can be determined by
\begin{equation}
{\mathcal B}_{\rm sig} = \frac{N_{\rm DT}}{N_{\rm ST}\cdot\epsilon_{\rm sig}},
\end{equation}
where $N_{\rm ST}$ and $N_{\rm DT}$ are the yields of the ST $\bar D^0$ mesons and the DT events in data, respectively;$\epsilon_{\rm sig}$ is the probability to reconstruct the signal under the condition that ST side was already reconstructed.

\subsection{ST selection}

The ST $\bar D^0$ candidates are selected with the same criteria as used in our previous works~\cite{epjc76,cpc40,bes3-pimuv,bes3-etaetapi,bes3-omegamuv,bes3-etamuv,bes3-etaX,bes3-DCS-Dp-K3pi,bes3-D-KKpipi,bes3-D-b1enu}.
For each charged track, the polar angle with respect to the MDC axis ($\theta$) is required to satisfy $|\cos\theta|<0.93$, and the point of closest approach to the interaction point must be within 1\,cm in the plane perpendicular to the MDC axis and within $\pm$10\,cm along the MDC axis.
Charged tracks are identified by using combined likelihoods from the $dE/dx$ and TOF measurements.  Tracks
are assigned as a pion (kaon) when that likelihood is larger than that for the kaon (pion) hypotheses.

Neutral pion candidates are reconstructed via $\pi^0\to\gamma\gamma$ decay, where the photon candidates are chosen from the EMC showers. The EMC time deviation from the event start time is required to be within [0,\,700]\,ns. The energy deposited in the EMC is required to be greater than 25\,(50)\,MeV if the crystal with the maximum deposited energy in that cluster is in the barrel~(end cap) region~\cite{BESCol}.
The opening angle between the photon candidate and the nearest charged track is required to be greater than $10^{\circ}$. For any $\pi^0$ candidate, the invariant mass of the photon pair is required to be within $(0.115,\,0.150)$\,GeV$/c^{2}$.
To improve the momentum resolution, a mass-constrained fit to the nominal $\pi^{0}$ mass~\cite{pdg2018} is imposed on the photon pair. The four-momentum of the $\pi^0$ candidate returned by this fit is used for further analysis.

In the selection of $\bar D^0\to K^+\pi^-$ events, the backgrounds from cosmic rays and Bhabha events are rejected by using the same requirements described in Ref.~\cite{deltakpi}.
To separate the ST $\bar D^0$ mesons from combinatorial backgrounds, we define the energy difference $\Delta E\equiv E_{\bar D^0}-E_{\mathrm{beam}}$ and the beam-constrained mass $M_{\rm BC}\equiv\sqrt{E_{\mathrm{beam}}^{2}/c^{4}-|\vec{p}_{\bar D^0}|^{2}/c^{2}}$, where $E_{\mathrm{beam}}$ is the beam energy, and $E_{\bar D^0}$ and $\vec{p}_{\bar D^0}$ are the total energy and momentum of the ST $\bar D^0$ meson candidate in the $e^+e^-$ CM frame.
If there is more than one $\bar D^0$ candidate combination in a specific tag mode, the one with the smallest $|\Delta E|$ is kept for further analysis.

\begin{figure*}[htbp]\centering
\includegraphics[width=1.0\linewidth]{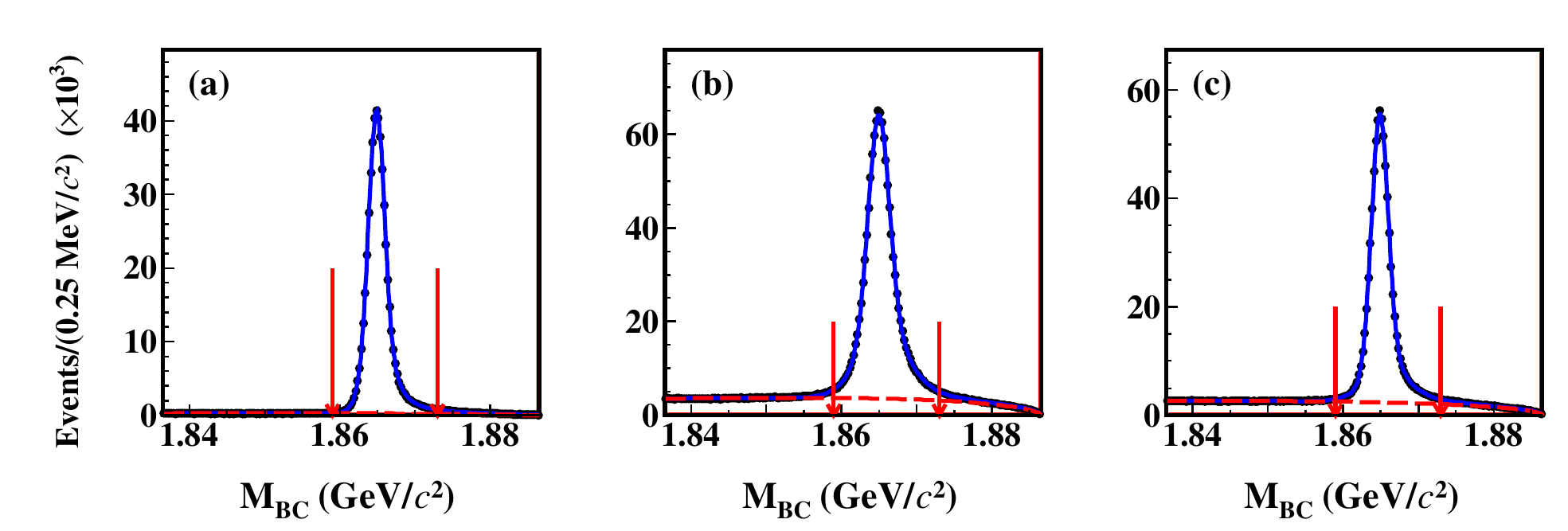}
\caption{
Fits to the $M_{\rm BC}$ distributions of the ST $\bar D^0$ candidates for (a) $\bar{D}^0\to K^+\pi^-$, (b) $\bar{D}^0\to K^+\pi^-\pi^0$, and (c) $\bar{D}^0\to K^+\pi^+\pi^-\pi^-$, respectively.
In each plot, the points with error bars are data, the red dashed curve is the background contribution, and the blue solid line shows the total fit.  Pairs of red arrows show the $M_{\rm BC}$ signal windows.}\label{fig:datafit_Massbc}
\end{figure*}

To suppress combinatorial backgrounds in the $M_{\rm BC}$ distributions, the ST $\bar D^0$ candidates are required to fall in $\Delta E\in (-55,40)$~MeV and $\Delta E\in (-25,25)$~MeV for the tag modes with and without a $\pi^0$ in the final states, respectively.
The $M_{\rm BC}$ distributions for various tag modes are shown in Fig.~\ref{fig:datafit_Massbc}. For each tag mode, the yield of the ST $\bar D^0$ mesons is obtained by a fit to the corresponding $M_{\rm BC}$ distribution. The signal is described by a probability density function (PDF) determined from the MC simulation (MC-determined PDF) convolved with a double-Gaussian function which describes the resolution difference between data and MC simulation. The background is parametrized by an ARGUS function~\cite{argus}. All parameters are left free in the fits.
Figure~\ref{fig:datafit_Massbc} shows the fit results to the $M_{\rm BC}$ distributions for individual ST modes. The candidates located in the $M_{\rm BC}$ signal region of $(1.859,1.873)$ GeV/$c^2$ are kept for further analysis. Summing over the three tag modes gives the total yield of the ST $\bar D^0$ mesons to be $2321009\pm1875$, where the uncertainty is calculated by the weighted average according to the fit results of the three tag modes.

\subsection{DT selection}

To avoid possible bias, a blind analysis technique is
followed where the data are analyzed only after the analysis
procedure is fixed and validated with MC simulation. The candidates for $D^{0}\to \bar p e^{+}$ and $D^{0}\to pe^{-}$ are selected from the remaining tracks and showers in the presence of
the tagged $D^0$ candidates. To obtain the information of $D^{0}\to \bar p e^{+}$ and $D^{0}\to pe^{-}$, we define $\Delta E^{\rm sig}$ and $M_{\rm BC}^{\rm sig}$ of the signal side similarly to those in the tag side.

Particle identification (PID) for electrons and positrons is performed by combining the $dE/dx$, TOF, and EMC measurements into confidence levels (CL) $CL_K$, $CL_\pi$, $CL_p$ and $CL_e$ for the kaon, pion, proton, and electron hypotheses.
Electron (positron) candidates are required to satisfy $CL_e>0.001$ and
\begin{equation}
\frac{CL_e}{CL_e+CL_\pi+CL_K}>0.8.
\end{equation}
To further suppress backgrounds due to misidentification between electrons (positrons) and hadrons, the ratio of the energy deposited in the EMC by the electron (positron) over its momentum~$(E/p)$ is required to be larger than 0.85$c$. To partially recover the effects of FSR and bremsstrahlung (FSR recovery), the four-momenta of clusters in the EMC within 10$^{\circ} $
of the initial positron direction are added to the positron four-momentum measured by the MDC.

The proton or anti-proton candidates are identified by using the $dE/dx$ and TOF measurements, from which combined confidence levels $CL_K$, $CL_\pi$, and $CL_p$ for the kaon, pion, and proton hypotheses are calculated, respectively.
The (anti-)proton candidates are required to satisfy $CL_p>0.001$, $CL_p>CL_K$, and $CL_p>CL_\pi$.

Studies of MC samples show that there remain a few backgrounds coming from mis-reconstructed proton candidates, \emph{e.g.}, $D^0\to K^- e^+ \nu_{\rm e}$ and processes other than $\psi (3770)\to D\bar D$. To suppress the background from $D^0\to K^- e^+ \nu_{\rm e}$, we define a variable of $U_{\rm miss}\equiv E_{\mathrm{miss}} -  |\vec{p}_{\mathrm{miss}}|\cdot c$,
where $E_{\mathrm{miss}}$ and $\vec{p}_{\mathrm{miss}}$ are the missing energy and momentum of the DT event in the $e^+e^-$ CM frame, respectively.
They are calculated by $E_{\mathrm{miss}}\equiv E_{\mathrm{beam}}-E_{K^{-}}-E_{e^{+}}$ and $\vec{p}_{\mathrm{miss}}\equiv\vec{p}_{D^0}-\vec{p}_{K^{-}}-\vec{p}_{e^{+}}$, where $E_{K^{-}(e^+)}$ and $\vec{p}_{K^{-}(e^+)}$ are the measured energy and momentum of the $K^-(e^+)$ candidates, respectively, and $\vec{p}_{D^0}\equiv-\hat{p}_{\bar D^0}\cdot \sqrt{E_{\mathrm{beam}}^{2}/c^{2}-m_{\bar D^0}^{2}\cdot c^{2} }$, where
$\hat{p}_{\bar D^0}$ is the unit vector in the momentum direction of the ST $\bar D^0$ meson and $m_{\bar D^0}$ is the nominal $\bar D^0$ mass~\cite{pdg2018}.
The use of the beam energy and the nominal $D^0$ mass for the magnitude of the ST $D^0$ meson momentum improves the $U_{\mathrm{miss}}$ resolution. For the correctly reconstructed events of $D^0\to K^{-} e^{+} \nu_{\rm e}$, the $U_{\mathrm{miss}}$ peaks around zero. The background from $D^0\to K^{-} e^{+} \nu_{\rm e}$ is suppressed by requiring $U_{\rm miss}$ to be outside the range of $(-0.15,0.15)$~GeV.

\subsection{Signal extraction}

\begin{figure*}[htbp]
\begin{center}
\subfigure[]{\includegraphics[width=0.45\linewidth]{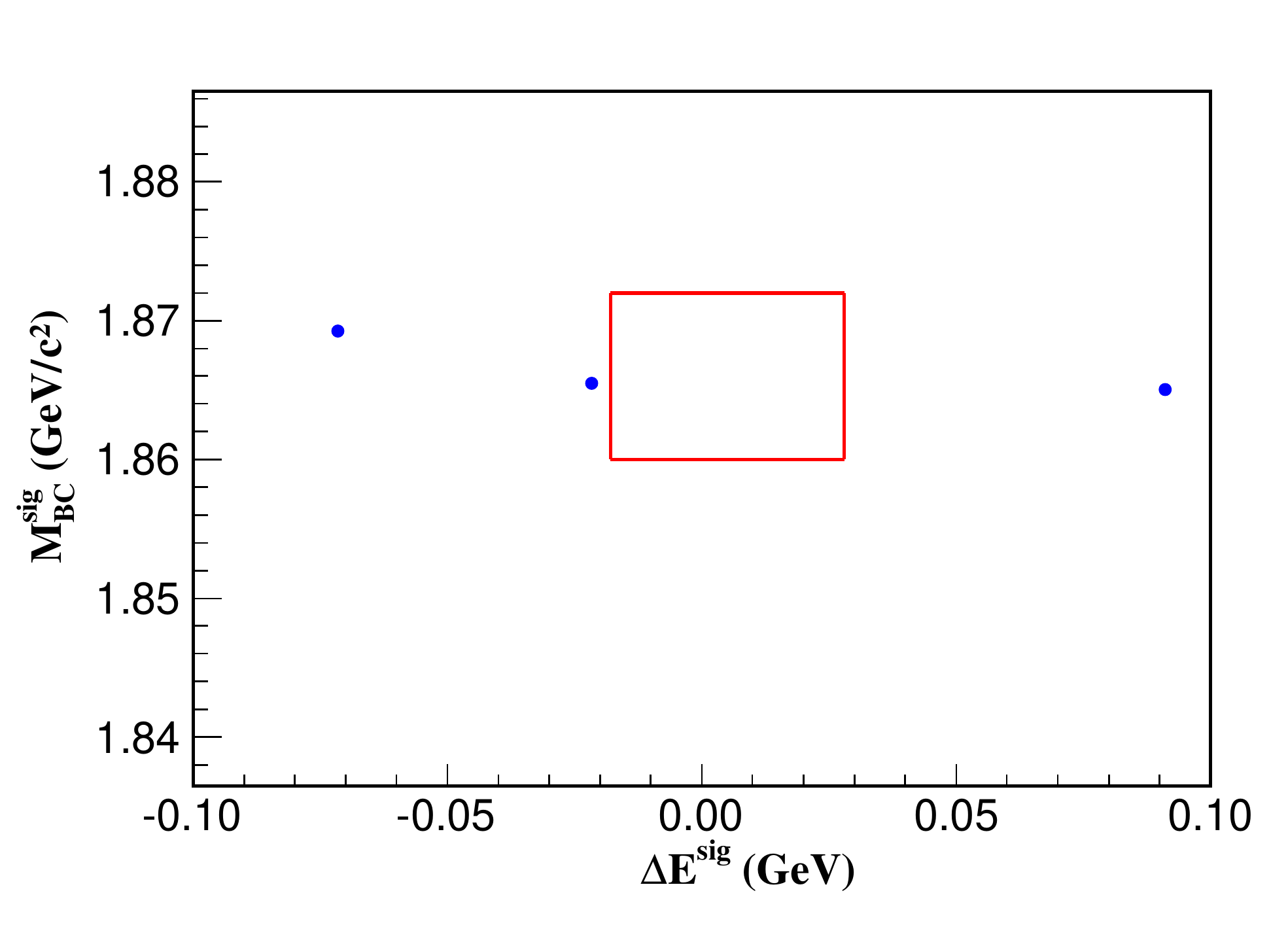}}
\subfigure[]{\includegraphics[width=0.45\linewidth]{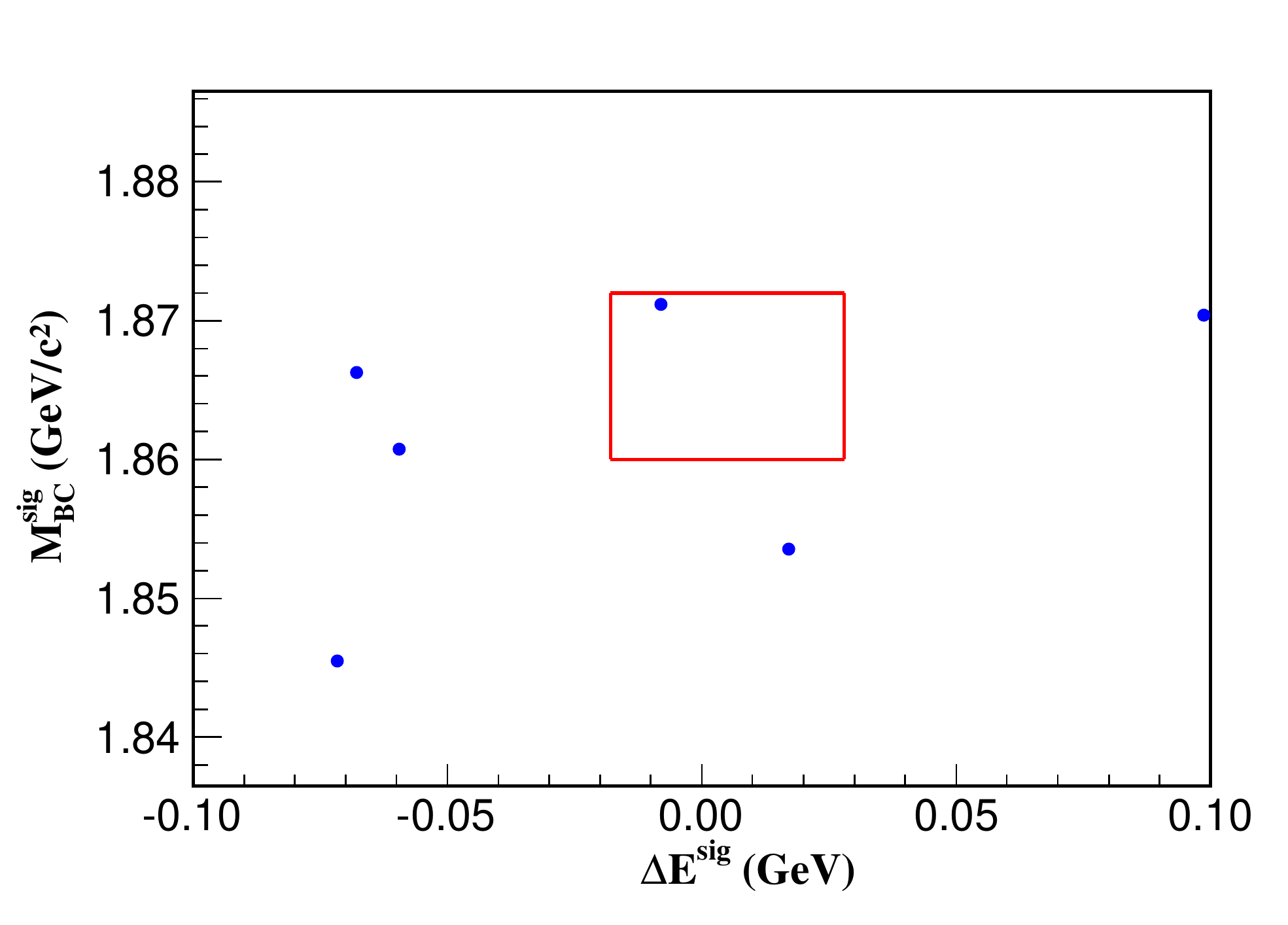}}
\label{fig:Uc}
\caption{Distributions of $M_{\rm BC}^{\rm sig}$ vs. $\Delta E^{\rm sig}$ of the candidate events for (a) $D^0\to \bar p e^+$ and (b) $D^0\to pe^-$ in data.
The red rectangles denote the signal region.
\label{fig:scatter}
}
\end{center}
\end{figure*}

Figures \ref{fig:scatter}(a) and \ref{fig:scatter}(b) show the distributions of $M_{\rm BC}^{\rm sig}$ vs. $\Delta E^{\rm sig}$ of the candidate events for $D^0\to \bar p e^+$ and $D^0\to pe^-$ selected from the data sample, respectively.
The signal yields are obtained by counting the events and conservatively assuming that the background events are evenly distributed. Since both of the $M_{\rm BC}^{\rm sig}$ and $\Delta E^{\rm sig}$ distributions from signal MC events have an asymmetric shape, to get a higher efficiency the signal regions are defined as $M_{\rm BC}^{\rm sig}~(-2.5\sigma_{M_{\rm BC}},4.0 \sigma_{M_{\rm BC}})$ vs.~$\Delta E^{\rm sig}~(-2.5\sigma_{\Delta E},2.0\sigma_{\Delta E})$, where $\sigma_{M_{\rm BC}}$ and $\sigma_{\Delta E}$ are the standard deviation of $M_{\rm BC}^{\rm sig}$ and $\Delta E^{\rm sig}$ which are obtained by fits to the signal MC.
The signal region is determined to be $1.860<M_{\rm BC}^{\rm sig}<1.872$~GeV/$c^2$ and $-0.028<\Delta E^{\rm sig}<0.018$ GeV for $D^{0}\to \bar{p}e^{+}(pe^{-})$.
We obtain the signal yields of the candidates for $D^0\to \bar p e^+$ and $D^0\to pe^-$ ($N^{\rm sig}$)
to be 0 and 1, respectively.

The background yields in the signal region are estimated by the events in
sideband region. In the whole region of $1.8365<M_{\rm BC}^{\rm sig}<1.8865$~GeV/$c^2$ and $-0.1<\Delta E^{\rm sig}<0.1$ GeV, we take the area outside of the signal region as our sideband region.
There are 3 and 5 background events in the sideband region ($N^{\rm BKG}$), for $D^0\to \bar p e^+$ and $D^0\to pe^-$, respectively.
The ratios of the signal region area over the sideband region area
($R_{\rm area}$) for $D^0\to \bar p e^+$ and $D^0\to pe^-$ are both 0.0587.
Multiplying $N^{\rm BKG}$ by $R_{\rm area}$ gives the expected background events in the signal region ($N^{\rm bkg}$)
to be 0.2 and 0.3 for $D^0\to \bar p e^+$ and $D^0\to pe^-$, respectively.

\subsection{DT efficiency}

To determine the detection efficiency, we simulate 100,000 events of $\bar{D}^0\to K^+\pi^-(\pi^0, \pi^+\pi^-)$ vs. $D^0\to \bar p e^+$ process and $\bar{D}^0\to K^+\pi^-(\pi^0, \pi^+\pi^-)$ vs. $D^0\to pe^-$ process for each tag mode, respectively, where $\bar{D}^0\to K^+\pi^-$ and the signal modes are modeled with a phase space generator, $\bar{D}^0\to K^+\pi^-\pi^0$ is modeled with the Dalitz~\cite{Kpipi0} generator and $\bar{D}^0\to K^+\pi^-\pi^+\pi^-$ are modeled with the partial wave analysis~\cite{K3pi} generator using measured distributions. The efficiencies of finding $D^0\to \bar p e^+$ or $D^0\to pe^-$ in the presence of the ST $\bar {D}^0$ meson ($\epsilon^{\rm sig}$)
are $(64.7\pm0.2)\%$ and $(64.9\pm0.2)\%$, respectively.

\section{Systematic uncertainty}

With the DT method, almost all systematic uncertainties related to the ST selection are cancelled and do not affect the BF measurement.
Table~\ref{sys} summarizes the remaining systematic uncertainties in the measurements of the BFs for
$D^0\to \bar p e^{+}$ and $D^0\to p e^{-}$. They are calculated relative to the measured BFs and are discussed below.

The systematic uncertainty of the total yield of the ST $\bar D^0$ mesons ($N_{\rm ST}^{\rm tot}$) is estimated to be 0.5\% \cite{epjc76,cpc40,bes3-pimuv}.

The systematic uncertainties of $e^\pm$ tracking and PID efficiencies are studied with a control sample of $e^+e^-\to\gamma e^+e^-$.
The difference of the $e^\pm$ tracking efficiencies between data and MC simulation, 1.0\%, is assigned as the systematic uncertainty of the $e^{\pm}$ tracking efficiency.
The systematic uncertainty from the $e^\pm$ PID efficiency is assigned to be 1.1\% per $e^\pm$.
Here, the obtained efficiencies in the control sample have been re-weighted to those in the signal decays in two dimensional (momentum and cos$\theta$) distributions.

The systematic uncertainties of proton tracking and PID efficiencies are studied using the control sample of $e^{+}e^{-}\to\pi^{+}\pi^{-}p\bar{p}$.
The systematic uncertainties of the proton tracking and PID efficiencies are assigned to be 1.0\% and 2.8\%, respectively.

To study the systematic uncertainties due to the signal region in $M_{\rm BC}^{\rm sig}$ and $\Delta  E^{\rm sig}$, we use the control sample of the DT candidate events for $D^0\to K^{-}K^{+}$.
The $M_{\rm BC}^{\rm sig}$ and $\Delta  E^{\rm sig}$ distributions of data are modeled with the MC-determined PDF convolved with a Gaussian resolution function.
After smearing the corresponding Gaussian resolution function for our signal MC events, the changes of the DT efficiencies
0.1\% and 0.3\% are taken as the systematic uncertainties of the $M_{\rm BC}^{\rm sig}$ and $\Delta  E^{\rm sig}$ window.

The uncertainty arising from limited MC statistics, 0.3\% for each signal decay mode, is considered as a source of systematic uncertainty.
The systematic uncertainty due to the FSR recovery is assigned to be 0.3\% by referring to that in large sample of $D^0\to K^-e^+\nu_e$~\cite{FSR2}.

We use the control sample of $D^0\to K^-e^+\nu_e$ to study the systematic uncertainties from $U_{\rm miss}$ requirement.
Since the efficiency differences caused by $U_{\rm miss}$ requirement between data and MC are very small, we ignore this term of systematic uncertainty.

Adding these systematic uncertainties in quadrature gives the total systematic uncertainties ($\Delta_{\rm syst}$) in the measurements of the BFs
for $D^0\to \bar p e^{+}$ or $D^0\to p e^{-}$ to be 3.5\%.

\begin{table}[htbp]
\centering
\caption{Relative systematic uncertainties (in \%) in the BF measurements.
}
%\resizebox{9.5cm}{4.5cm}{
\begin{tabular}{lcc}
  \hline
  \hline
  Decay                              &$D^0\to \bar p e^{+}$   &$D^0\to p e^{-}$  \\
  \hline
  $N^{\rm tot}_{\rm ST}$             &0.5       &0.5      \\
  $e^\pm$ tracking                   &1.0       &1.0      \\
  $e^\pm$ PID                        &1.1       &1.1      \\
  $p(\bar p)$ tracking               &1.0       &1.0      \\
  $p(\bar p)$ PID                    &2.8       &2.8      \\
  $M_{\rm BC}^{\rm sig}$ requirement &0.1       &0.1      \\
  $\Delta E^{\rm sig}$ requirement   &0.3       &0.3      \\
  MC statistics                      &0.3       &0.3      \\
  FSR recovery                       &0.3       &0.3      \\
  \hline
  Total  ($\Delta_{\rm syst}$)       &3.5       &3.5      \\

  \hline
  \hline
\end{tabular}
\label{sys}
\end{table}

\section{Results}

The ULs on the numbers of signal events at 90\% CL are calculated by using a frequentist
method~\cite{frequent} with unbounded profile likelihood treatment
of systematic uncertainties, as implemented by the TROLKE package in the ROOT software~\cite{ROOT}, with the numbers of $N^{\rm sig}$, $N^{\rm bkg}$, $\epsilon^{\rm sig}$, and $\Delta_{\rm syst}$ documented above.
Here, the numbers of the signal and background events are assumed to follow a Poisson distribution,
and the detection efficiency is assumed to follow a Gaussian distribution.

The ULs on the BFs are calculated to be
\begin{eqnarray}
{\mathcal B}_{D^0\to \bar{p} e^+} < 1.2\times 10^{-6} \nonumber
\end{eqnarray}
and
\begin{eqnarray}
{\mathcal B}_{D^0\to p e^-} < 2.2\times 10^{-6}, \nonumber
\end{eqnarray}
respectively.

\section{Summary}

In summary, by analyzing an $e^+e^-$ annihilation data sample corresponding to an integrated luminosity of 2.93~fb$^{-1}$ collected with the BESIII detector, we have searched for the SM forbidden decays $D^0\to \bar p e^+$ and $D^0\to p e^-$. No obvious signals have been observed. The ULs on $\mathcal{B}(D^0\to \bar p e^+)$ and $\mathcal{B}(D^0\to p e^-)$ at $90\%$ CL are set to be $1.2\times10^{-6}$ and $2.2\times10^{-6}$, respectively.
Our ULs are the most stringent ones to date for these processes, but are still far above the prediction of the higher generation model~\cite{su5,pre}.

\section{Acknowledgement}

The BESIII collaboration thanks the staff of BEPCII and the IHEP computing center for their strong support. This work is supported in part by National Key R\&D Program of China under Grants No. 2020YFA0406400 and No. 2020YFA0406300; National Natural Science Foundation of China (NSFC) under Grants No. 12035009, No. 11875170, No. 11475090, No. 11625523, No. 11635010, No. 11735014, No. 11822506, No. 11835012, No. 11935015, No. 11935016, No. 11935018, No. 11961141012, No. 12022510, No. 12025502, No. 12035013, No. 12061131003, No. 12192260, No. 12192261, No. 12192262, No. 12192263, No. 12192264, and No. 12192265; the Chinese Academy of Sciences (CAS) Large-Scale Scientific Facility Program; Joint Large-Scale Scientific Facility Funds of the NSFC and CAS under Grants Nos. U1732263, U1832207; CAS Key Research Program of Frontier Sciences under Grant No. QYZDJ-SSW-SLH040; 100 Talents Program of CAS; INPAC and Shanghai Key Laboratory for Particle Physics and Cosmology; ERC under Grant No. 758462; European Union Horizon 2020 research and innovation programme under Grant No. Marie Sklodowska-Curie grant agreement No 894790; German Research Foundation DFG under Grants Nos. 443159800, Collaborative Research Center CRC 1044, FOR 2359, GRK 214; Istituto Nazionale di Fisica Nucleare, Italy; Ministry of Development of Turkey under Grant No. DPT2006K-120470; National Science and Technology fund; Olle Engkvist Foundation under Grant No. 200-0605; STFC (United Kingdom); The Knut and Alice Wallenberg Foundation (Sweden) under Grant No. 2016.0157; The Royal Society, UK under Grants Nos. DH140054, DH160214; The Swedish Research Council; U. S. Department of Energy under Grants Nos. DE-FG02-05ER41374, DE-SC-0012069.

\end{document}